\documentclass[fleqn,usenatbib]{mnras}

\usepackage{newtxtext,newtxmath}

\usepackage[T1]{fontenc}

\DeclareRobustCommand{\VAN}[3]{#2}
\let\VANthebibliography\thebibliography
\def\thebibliography{\DeclareRobustCommand{\VAN}[3]{##3}\VANthebibliography}

\usepackage{graphicx}	
\usepackage{amsmath}	

\usepackage{ulem}

 \renewcommand{\(}{\left(}
 \renewcommand{\)}{\right)}

\renewcommand{\v}[1]{\boldsymbol{#1}}

\newcommand{\mdot}{\dot{M}}
\newcommand{\Msun}{\,M_{\odot}}	
	
\newcommand{\vinf}{\varv_{\infty}}	
\newcommand{\kms}{\,~\textrm{km\,s}^{-1}}	
\newcommand{\teff}{T_{\mathrm{eff}}}

\title[Hydrodynamic simulation of Cygnus OB2]{Hydrodynamic simulation of Cygnus OB2: the absence of a cluster wind termination shock}

\author[T. Vieu et al.]{
T. Vieu,$^{1}$\thanks{E-mail: thibault@mpi-hd.mpg.de}
C. J. K. Larkin,$^{1,2,3}$
L. H\"arer,$^{1}$
B. Reville,$^{1}$
A.A.C. Sander,$^{2}$
V. Ramachandran$^{2}$
\\
$^{1}$Max-Planck-Institut f\"ur Kernphysik, Saupfercheckweg 1, D-69117 Heidelberg, Germany \\
$^{2}$Zentrum f\"{u}r Astronomie der Universit\"{a}t Heidelberg, Astronomisches Rechen-Institut, M\"{o}nchhofstr. 12-14, 69120 Heidelberg, Germany \\
$^{3}$Max-Planck-Institut f\"{u}r Astronomie, K\"{o}nigstuhl 17, D-69117 Heidelberg, Germany\\
}

\date{Accepted XXX. Received YYY; in original form ZZZ}

\pubyear{2024}

\begin{document}
\label{firstpage}
\pagerange{\pageref{firstpage}--\pageref{lastpage}}
\maketitle

\begin{abstract}
We perform a large-scale hydrodynamic simulation of a massive star cluster whose stellar population mimics that of the Cygnus OB2 association. The main-sequence stars are first simulated during 1.6~Myr, until a quasi-stationary state is reached. At this time the three Wolf-Rayet stars observed in Cygnus OB2 are added to the simulation, which continues to 2~Myr. Using a high-resolution grid in the centre of the domain, we can resolve the most massive stars individually, which allows us to probe the kinetic structures at small (parsec) scales. We find that, although the cluster excavates a spherical ``superbubble'' cavity, the stellar population is too loosely distributed to blow a large-scale cluster wind termination shock, and that collective effects from wind-wind interactions are much less efficient than usually assumed. This challenges our understanding of the ultra-high energy emission observed from the region. 
\end{abstract}




\section{Introduction}
The Cygnus X complex is a nearby region of ongoing star-formation at a distance of about 1.6~kpc. It hosts a number of OB associations, including Cygnus OB2, a young massive star cluster containing thousands of OB stars \citep{massey1991,knodlseder2000} among which tens are known to be powerful early O-stars \citet{Wright2015,Berlanas2020}. The strong mechanical feedback imparted by the cluster is expected to shape the surrounding environment and impact stellar formation. This in turn is anticipated to influence non-thermal particle acceleration and emission from the cluster. However, despite decades of observational and theoretical investigations, there is still a limited understanding of the physical processes that account for observations of the region. One of the chief difficulties is the presence of several foreground structures which hinder the interpretation of multi-wavelength observations \citep{Uyaniker2001,zhang2024} and the reconstruction of the gas distribution. Radio and X-ray observations hint at the presence of a large-scale bubble \citep[e.g.,][]{albacete2023}, most likely heated by the stellar feedback, although the boundaries do not seem to coincide with the dense ``shell'' structure that is theoretically expected \citep{weaver1977}.

The Cygnus X region has become a source of great interest in the field of high-energy astrophysics since the discovery of coincident diffuse gamma-ray emission first measured by the Fermi satellite \citep{ackermann2011}, and subsequently detected by the HAWC and LHAASO observatories \citep{gammaCygnus_HAWC2021,cao2021}. In particular, the LHAASO collaboration reported the detection of gamma-rays above 1~PeV, which suggests that the Cygnus region hosts a hadronic ``PeVatron'' \citep{lhaasocygnus2024}. Identifying and understanding acceleration sites within the Cygnus~X region could unveil a new class of cosmic-ray sources which, given the growing number of observational constraints challenging the classical supernova paradigm \citep[e.g.,][]{gabici2019}, is key to understanding the origin of galactic cosmic rays at very-high energies.

Star-forming regions are intricate environments with non-trivial feedback amongst individual components. The massive stars blow powerful supersonic winds which create, in the early-stage of the cluster evolution, a collection of expanding cavities called \textit{wind-blown bubbles} \citep{weaver1977}. After a few tens of kyrs, the wind-blown bubbles percolate to form a large-scale \textit{superbubble} filled with a hot, high pressure and rarefied plasma. Inside the superbubble, each massive star continues to blow a supersonic wind, which terminates at a distance of approximately 1~pc away from the star, at its \textit{stellar wind termination shock}. If the cluster is compact enough (typically if its half-mass radius is of the order of 1~pc), the stellar winds interact so efficiently that they are expected to merge into a single, large-scale ($10-20$~pc) supersonic outflow, which terminates at a so-called \textit{cluster wind termination shock}. On the other hand, if the stars are too loosely distributed, the interactions between the supersonic winds might not be enough to drive the formation of such a large-scale cluster wind termination shock \citep{gupta2020}. In the end, the environment in and around a stellar cluster might contain several strong shocks from the interactions between supersonic outflows, which is also expected to generate a high level of turbulence \citep{vieu2024core}. This makes these regions favourable sites for particle acceleration.

Several scenarios of efficient particle acceleration in star-forming regions have been proposed in the past, including re-acceleration processes in the core \citep[e.g.][]{bykov1992b,klepach2000,bykov2013}, re-acceleration processes in the low-density superbubble \citep[e.g.][]{ferrand2010,vieu2022}, acceleration at the cluster wind termination shock \citep[e.g.][]{morlino2021}, acceleration by a supernova remnant expanding in the close vicinity of the cluster core \citep{vieu2023}. However, most of these scenarios fail to accelerate PeV protons \citep{vieu2022Emax}, which are necessary to explain the LHAASO data. Shortcomings of the re-acceleration processes in the core have been especially addressed in recent work \citep{vieu2024core}, where it was shown that re-acceleration processes become inefficient in expected cluster-core conditions. It seems therefore that a scenario such as the one investigated by \citet{bykov2022}, where particles are re-accelerated in an ensemble of stochastic shocks interacting in a very compact cluster core, is excluded for Cygnus~OB2. The presence of turbulence in the low density cavity beyond the cluster core might enhance the acceleration processes, although this is not expected to produce PeV protons \citep{vieu2022Emax}, especially in the case of Cygnus~OB2 which, with an estimated mechanical power of a few $10^{38}$~erg\,s$^{-1}$, is not the most powerful stellar cluster in our Galaxy.

Another possible scenario to account for the gamma-ray observations is that of efficient particle acceleration at a large-scale \textit{cluster wind termination shock} (WTS). \citet{menchiari2024} recently explored this possibility and argued that it could provide a satisfactory explanation to the data, although the authors disregarded the latest LHAASO measurements. This model seems promising, however it entirely relies on the hypothesis that Cygnus~OB2 is able to blow a large-scale cluster WTS, which is as yet unclear.

With an estimated power of $2-3 \times 10^{38}$~erg\,s$^{-1}$ and a age of a few Myr \citep{Wright2015}, the one-dimensional hydrodynamic theory of \citet{weaver1977} predicts a WTS size of about 13~pc, which could in principle be enough to accelerate PeV protons if Bohm diffusion is assumed \citep{morlino2021}. However, Cygnus OB2 is a rather peculiar stellar cluster, closer to a loose association of massive stars than a young compact cluster such as Westerlund~1. Indeed, the stellar power is not concentrated in a compact region, but distributed within about 15~pc (in particular, the three Wolf-Rayet stars, which dominate the mechanical output, are strongly off-centred). In this case, the putative cluster WTS is \textit{a minima} weak. Using the calculations by \citet{canto2000}, one finds a sonic Mach number of 2.6, which, in a textbook diffusive shock acceleration scenario, would produce a spectrum much steeper ($f(p) \sim p^{-4.7}$) than what is required to explain the gamma-ray observations ($f(p) \sim p^{-4.2}$ according to \citealt{menchiari2024}). On the other hand, it is not clear if there is at all such a large-scale collective termination shock around Cygnus OB2. It is usually believed that a collective wind can form if and only if the cluster WTS predicted by \citet{weaver1977} extends beyond the extension of the cluster core. While this is surely a necessary condition, it may not be sufficient. In particular, if the distance between the O stars is larger than the individual stellar WTSs blown over a few Myr, one should not expect efficient wind-wind interactions, and the notion of a \textit{collective wind} becomes ill-defined.

Beyond phenomenological arguments, detailed simulations are key to properly understand star-forming regions, in particular to probe the properties of the plasma, understand the kinetic structures and identify the sites of efficient particle acceleration. There have been many efforts in the recent years towards global hydrodynamic or magnetohydrodynamic simulations of stellar clusters forming superbubbles, often aiming at probing the feedback of the superbubble onto the interstellar medium at very large (100s~pc) scales. For this purpose, the star clusters are often modelled as point-like injections of energy, and the impact of supernovae is studied in detail \citep[e.g.][]{agertz2013,krause2013,kim2017,elbadry2019,lancaster2021}. These studies provide information on the expansion of the superbubble shell, the impact of thermal conduction, radiative cooling, photoionisation, density fluctuations, the occurrence of shell instabilities, or the impact of the presence of clumps and turbulence in the vicinity of the cluster. This is relevant not only to understand how the stellar feedback carves the parent molecular cloud, but also to investigate the dynamics of the Galaxy at large scales, e.g., the distribution of hot gas and the regulation of star-formation \citep[e.g.][]{gatto2017}. However, in order to investigate structures at intermediate scales ($\sim 10$~pc), refined simulations must be set up, which is particularly relevant when the cluster wind and supernovae are blown inside an inhomogeneous, dense, and turbulent molecular cloud \citep[e.g.][]{rogers2013,geen2021}. In these simulations, the feedback is found to be highly asymmetric and deviates strongly from the theoretical picture, although the cluster is still modelled as a point-like source.

In order to understand the structure of the supersonic flow in the close vicinity of the cluster, and therefore the shape of the cluster wind termination shock, one needs to go beyond a point-like injection of energy. A possible improvement is to inject each individual star as a source of thermal energy \citep{gupta2018}, which effectively creates an extended, continuous, injection region. Although this might be a good approximation to model the feedback of a very compact and powerful star cluster -- such as Westerlund 1 where several hundreds of OB stars are gathered within a core of 1~pc \citep{clark2005} -- it is less relevant in the case of a more extended cluster like Cygnus OB2, where approximately 80 O stars are distributed over about 15~pc. Besides, as will be discussed in detail in Section~\ref{sec:obsCygOB2}, the wind power in Cygnus OB2 is dominated by a few very powerful O and Wolf-Rayet stars. In such a case, it is necessary to simulate each star individually using a resolution that is high enough to resolve the regions of wind-wind interactions at sub-parsec scales \citep{badmaev2022,vieu2024core}. It is computationally challenging to resolve such small scales while keeping track of the medium-size structures ($10-20$~pc) inside a numerical box that is large enough to contain the large-scale superbubble ($50-100$~pc) over millions of years.

The goal of the present work is two-fold: Firstly, we aim at highlighting astrometric results which are often oversimplified by the high-energy astrophysics community. Secondly, we probe the presence of shocks inside the superbubble by running a high-resolution 3D hydrodynamic simulation of a cluster whose distribution of massive stars statistically matches that of Cygnus OB2.

Section~\ref{sec:obsCygOB2} is devoted to a detailed description of the massive star population of Cygnus~OB2 which is then used as a basis to set-up our simulation. Section~\ref{sec:results} described the results of the simulation, highlighting in particular the absence of a large-scale cluster WTS around the cluster, and discussing consequences on scenarios of particle acceleration. We conclude in Section~\ref{sec:conclusion}.

\section{Simulating the massive star population in Cygnus~OB2}\label{sec:obsCygOB2}
The aim of this section is to model Cygnus~OB2 as accurately as allowed by the available data. In order to get around uncertainties (e.g. in parallaxes measurements or in stellar properties), we will have to make assumptions and extrapolations. These will be systematically designed to maximise the mechanical feedback of the cluster, i.e. our model cluster will be likely more compact and more powerful than the real configuration of Cygnus~OB2.

\begin{figure}
          \centering
              \includegraphics[width=\linewidth]{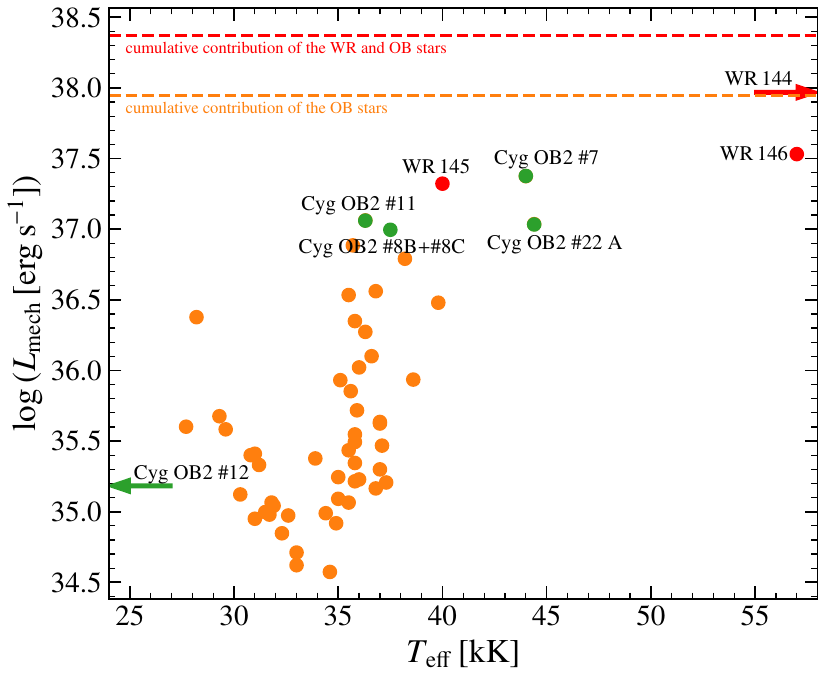}
   \caption{Mechanical energy versus effective temperature for the considered OB and WR stars in Cygnus OB2. Orange dots are denoting the O stars with known wind strength parameters from \citet{Berlanas2020}. Specific OB and WR stars are outlined in green or red respectively. For objects with effective temperatures outside the plot range, arrows are given instead of dots. The dashed orange line shows the integrated mechanical luminosity of all O stars with individually known wind strength parameters. The red dashed line shows the total mechanical luminosity including the three WR stars. }
   \label{fig:emech-teff}
 \end{figure}

\subsection{Stellar population from observations}

Cygnus OB2 is one of the best examples of a young massive star cluster close to Earth, with a total stellar mass estimated around $16 \, 500^{+3800}_{-2800}$~M$_\odot$ \citep{Wright2015}. Its massive star population has been continuously studied and extended over the years \citep[e.g.,][]{Comeron2002,Comeron2012,Wright2015,Berlanas2018}, although the high extinction in its direction \citep[$A_V \sim 5 - 6\,$mag,][]{Wright2015} hinders the observations. The OB population is still not considered to be complete, nevertheless this mainly limits the completeness with respect to faint, late-type O stars which are not expected to provide a significant contribution to the total mechanical luminosity of the cluster \citep[][]{Berlanas2018}.

Cygnus~OB2 is located in the northern sky, at a distance closer than 2\,kpc \citep{Rygl2012}. The Gaia parallaxes reveal substructures along the line of sight \citep{Berlanas2019,Orellana2021}, namely a small complex in the foreground at $\sim$$1.4\,$kpc and a main complex at $\sim$$1.7\,$kpc. The latter is further divided into two groups of stars centred around the massive systems Cygnus~OB2~\#22 and Cygnus~OB2~\#8 \citep{Bica2003}, which are sometimes referred to as clusters themselves named ``Bica 1'' and ``Bica 2'' \citep[e.g.,][]{MaizApellaniz2020}. Both systems contain several luminous, early-type O stars including one of the rare O3 spectral class in each of them, namely Cygnus~OB2 \#7 
(O3f*) in Bica 2 and Cygnus~OB2 \#22 A (also O3f*) in Bica 1 \citep{Sota2011}. The two clusters form a double-cluster as they are located only $\sim$$0.1\,$deg away from each other in the projected plane on the sky.

In the present work, we consider the area within $1\deg^2$ centred on Cygnus OB2 \#8 (i.e., the centre of the ``Bica 2 cluster'') as in e.g. \citet{Wright2015}. We assume a cluster distance of $1.65\,$kpc, motivated by the recent analysis from \citet{MaizApellaniz2020}, who obtained a Gaia-based distance of $1.64^{+0.13}_{-0.11}\,$kpc considering the Bica 2 group. For comparison, the derived distance by \citet{MaizApellaniz2020} for Bica 1 (containing Cyg OB \#22) is $1.72^{+0.14}_{-0.12}\,$kpc.

The age of Cygnus~OB2 is ill-defined as the region is not coeval. Estimates range between $2 \pm 1$ Myr for a population of mid-type O and B supergiant (BSG) stars \citep{Hanson2003} up to 5--7~Myr for a population of A-type stars \citep{Drew2008}. Star-formation has likely happened over the last 6 Myr, with two main starburst events at approximately 3 and 5 Myr \citep{Berlanas2020}. It is believed that the current OB stellar population formed likely later than most of the observed lower-mass stars, which has also been concluded for other OB clusters \citep[e.g.,][]{Povich2010, Ramachandran2018, Schneider2018}.

Three Wolf-Rayet (WR) stars can be found in the vicinity of Cygnus~OB2. These stars provide a significant fraction of the total stellar wind power, and need to be considered in detail due to their positions far from the centre of the cluster. WR 144 is a WC4-type star with $\log \mdot = -4.62$ and $\vinf = 3500 \kms$ \citep{sander2019}. WR 145 is a WR+O7V binary with a transition type (WN7/WC) primary \citep{Muntean2009}. We cannot resolve the individual stars in the system and thus use the values of $\log \mdot = -4.49$ and $\vinf = 1440 \kms$ from the single-star analysis by \citet{Sander2012,sander2019} which provides a conservative estimate of the mass-loss rate. WR 146 is a WC6+O8I binary, with $\vinf = 2900 \kms$ \citep{Eenens1994} for the WR component, though the mass-loss rate is uncertain, both for the individual components and for the combined spectrum. We therefore make an assumption based on the average value for Galactic WC stars from Table 5 in \citet{sander2019} and scale this to the luminosity of WR 146, giving us $\log \mdot \approx -4.9$ \citep[see, e.g.,][for the corresponding equations]{sander2019}. This estimate is compatible with recent analyses showing that the mass-loss rate of WR146 is likely smaller than previously thought \citep{zhekov2017,pittard2021WR146}.

In addition to the WR stars, we also specifically model Cygnus~OB2 \#12, a blue hypergiant of spectral type B3Ia$^+$ and one of the most luminous star in the Milky Way \citep{Clark2012}. The star is possibly a Colliding Wind Binary \citep{Oskinova2017}, but similar to other binary systems the components cannot be treated individually. We therefore adopt $\vinf = 400 \kms$ and $\log \mdot = -5.52$ from \citet{Clark2012}. 

\subsection{Simulated population of massive stars} \label{subsec:simulatedpopulation}
The main data source for the stars in our simulation is the spectroscopic sample of 78 O stars from \citet{Berlanas2020}, which marks the most complete spectral sample for the O-star population of Cygnus~OB2 so far. 
Using a grid of FASTWIND \citep{Puls2005,RiveroGonzalez2012} stellar atmosphere models and the \textsc{iacob-gbat} tool \citep{SimonDiaz2011,Holgado2018}, \citet{Berlanas2020} have provided the effective temperature $\teff$, the stellar radius $R_\star$ and the wind-strength parameter
\begin{align}
    Q &= \frac{\dot{M}}{M_\odot/\mathrm{yr}} \left( \frac{\kms}{\varv_\infty} \frac{R_\odot}{R_{\star}} \right)^{3/2}
    \, ,
    \label{eq:windstrength}
\end{align}
for the 52 O stars in Cygnus~OB2 that have reliable Gaia DR2 astrometry. Given the considerable reddening of Cygnus~OB2, the analysis of \citet{Berlanas2020} is limited to the optical wavelength range. Without additional constraints, e.g., UV measurements of the terminal wind velocity, $\varv_\infty$, the combined wind strength parameter $Q$ can be robustly constrained from the spectral fit but not the individual values of $\varv_\infty$ and the mass-loss rate $\dot{M}$. As we wish to know the mechanical luminosity 
\begin{equation}
  \label{eq:lmech}
  L_\text{mech} = \frac{1}{2} \dot{M} \varv_\infty^2
  \, ,
\end{equation}
for our simulation, we need an additional estimate for $\varv_\infty$ to calculate $\dot{M}$ via Eq.\,\eqref{eq:windstrength}. In principle, we could estimate the terminal wind velocity from the (effective) escape velocity using the modified-CAK relation between them \citep{castor1975,KudritzkiPuls2000}, but given the considerable uncertainties of spectroscopic masses and the available quantities from \citet{Berlanas2020}, we instead use the empirically well-constrained relation between $\varv_\infty$ and $\teff$. The recent determination by \citet{Hawcroft2023} for O stars at Galactic metallicity yields the formula
\begin{align}
    \vinf &= \left( 0.102 \frac{\teff}{\mathrm{K}} - 1300 \right) \kms \, ,
    \label{eq:hawcroft}
\end{align}
which we use to estimate values of $\vinf$ and thus obtain both $\dot{M}$ and $L_\text{mech}$, avoiding any explicit assumptions about the spectroscopic mass. We note that the values for $Q$ derived by \citet{Berlanas2020} and also the inferred $\dot{M}$ values assume smooth winds and thus the real mass-loss rates could even be a bit lower \citep[see][for recent reviews]{Puls2008,Hamann2008,Oskinova2016}. 
Moreover, Eq.\,\eqref{eq:hawcroft} only reproduces a statistical trend with a non-negligible spread, meaning that the values for individual objects could be off (see Table~\ref{tab:powerstellarpopulation}), but given the aims of our simulations, these are expected to average out. In any case, as we will discuss further below, we do not aim at an exact reproduction of Cygnus~OB2, but instead a statistically equivalent cluster.

There remain 26 O stars for which the stellar radius or wind parameters are not known. In order to complete our sample, we generate a synthetic stellar population by exactly sampling the initial mass function of Cygnus OB2, with a slope of $\Gamma = 1.39$ \citep{Wright2015}. We pick stars with masses between 20 and 50 $\Msun$ in order to avoid introducing unusually powerful stars which are not seen in the observations. On the other hand, starting at 20 $\Msun$ is expected to produce more powerful stars than in reality, since the remaining sample is likely biased toward lower masses. Overestimating the power provides an optimistic case for the formation of a cluster WTS around Cygnus OB2.

To infer plausible values for the wind parameters from the initial mass $M$, we use the prescriptions of \citet{seo2018} (in the case of non-rotating single stars): 
\begin{align}
    &\log_{10} \frac{\dot{M}}{{ M_\odot\,\mathrm{yr}^{-1}}}
    = -3.38 \left( \log_{10}  \frac{M}{M_\odot} \right)^2 + 14.59 \log_{10} \frac{M}{M_\odot} - 20.84 \, ,
    \label{stellarparametersMdot}
    \\
    &\log_{10} \frac{\vinf}{\textrm{km\,s}^{-1}} = 0.08 \log_{10} \frac{M}{M_\odot} +3.28 \, .
    \label{stellarparametersVw}
\end{align}

In the end we obtain a population of O stars with a mechanical power of $10^{38}$~erg/s. The mechanical luminosities of this OB stellar population is shown in Fig.~\ref{fig:emech-teff}, including their total sum. When adding the three nearby WR stars, as well as the Blue Hypergiant Cygnus~OB2 \#12, it becomes evident that the latter, despite its high luminosity, has very little impact on the overall budget. This is due to the comparably low terminal velocity of the B3-4 hypergiant of only $400\,$km/s. 
When including the WR stars (cf.\ Fig.\,\ref{fig:emech-teff}), we obtain a total mechanical power of $2.6 \times 10^{38}$~erg/s for the Cygnus OB2 cluster. The most powerful stars are listed in Table~\ref{tab:powerstellarpopulation}. Note that the whole O star population represents only about 40\% of the total mechanical power in the cluster with the remaining 60\% being generated by the three WR stars. This underlines the importance to properly identify WR stars in stellar populations as they can completely change the mechanical energy budget.

We also show estimated uncertainties in Table~\ref{tab:powerstellarpopulation}. These are derived by standard error propagation, assuming conservative standard values for $\epsilon(\log \mdot) = 0.15 \text{ [dex]}$ and $\epsilon(\vinf) = 200 \kms$.

\begin{table*}
\centering
\begin{tabular}{lccccccccc}
\hline
 & WR144 & WR145 & WR146 & Cygnus~OB2 \#7 & Cygnus~OB2 & Cygnus~OB2 \#11 & Cygnus~OB2 \#22 &  O stars & O stars \\ 
 &   &   &   &   & \#8B+\#8C  &   &   &  (total) & (simulated) \\
 \hline
$L_w$ & $9.3 \pm 2.7$ & $2.1 \pm 0.95$ & $3.4 \pm 1.0$ & $2.4 \pm 0.71$ & $0.99 \pm 0.32$  & $1.2 \pm 0.39$   & $1.1 \pm 0.32$ & $10$ & $7.7$  \\ 
\hline
\end{tabular}
\caption{Distribution of the stellar power in Cygnus OB2, in units of $10^{37}$~erg/s. Most of the feedback is blown by the WR stars and 5 very powerful O stars. The discrepancy between the total sample and the simulated sample is due to the stars located beyond 12~pc being discarded because of numerical limitations, as discussed in Section~\ref{subsec:3Dpositions}.}
\label{tab:powerstellarpopulation} 
\end{table*}

Although the simulated stars are not individually identical to that of Cygnus OB2, our procedure allows to generate a synthetic cluster that is statistically equivalent to the real distribution. In any case, the precise power of a standard O star is not expected to strongly affect the simulation results. 

\subsection{3D positioning}\label{subsec:3Dpositions}
\begin{figure}
    \centering
    \includegraphics[width=1\linewidth]{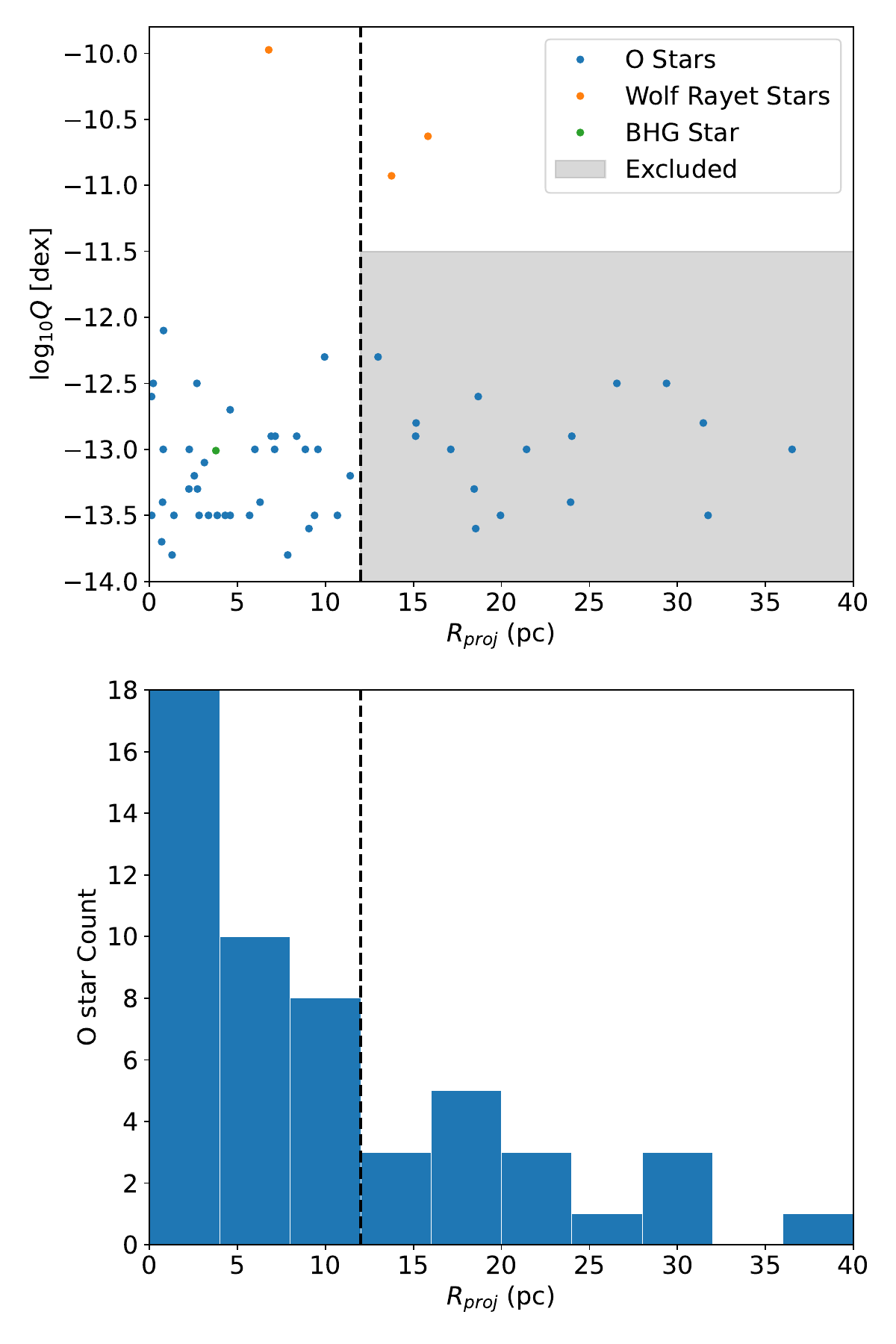}
    \caption{Statistical properties of our sample. The black dashed line at 12 pc in each panel shows where we cut off our simulated population. The upper panel shows $\log_{10} Q$ as a function of projected radial distance $R_\mathrm{proj}$ for the individual stars where we have $Q$. The lower panel is a histogram showing the number of O stars with $R_\star$ and $Q$ available.}
    \label{fig:astrometry}
\end{figure}

The most important factor that influences the feedback is the spatial distribution of the stars. Indeed, one should expect qualitatively different results if all the stars are gathered in a compact core, or rather loosely scattered in a large volume. The radial distribution of Cygnus~OB2 as catalogued in \citet{Berlanas2020} is shown in Fig.~\ref{fig:astrometry}. The histogram in the bottom panel shows that Cygnus~OB2 is actually neither very compact nor very loose. Although most of the stars and most of the wind power are concentrated in the inner few parsecs, the distribution does extend up to large distances. In particular, the WR stars are notably off-centred, at projected distances of 6.8~pc (WR144), 13.8~pc (WR146) and 15.8~pc (WR145). In fact, most of the cluster mechanical power is blown by a few stars, as seen in Fig.~\ref{fig:emech-teff} and summarised in Table~\ref{tab:powerstellarpopulation}: the three WR stars, the very powerful O star Cygnus~OB2~\#7 , and eventually the stars Cygnus~OB2~\#8B and Cygnus~OB2~\#8C, which are in such close vicinity that they can be considered as a single star from the point of view of their feedback. It is clear that such a scattered distribution of stars cannot be approximated by a continuous distribution. Massive stars need to be resolved individually in order to properly encapsulate the wind-wind interactions and a 3D positioning is required for all the simulated stars.

\begin{figure}
          \centering
              \includegraphics[width=\linewidth]{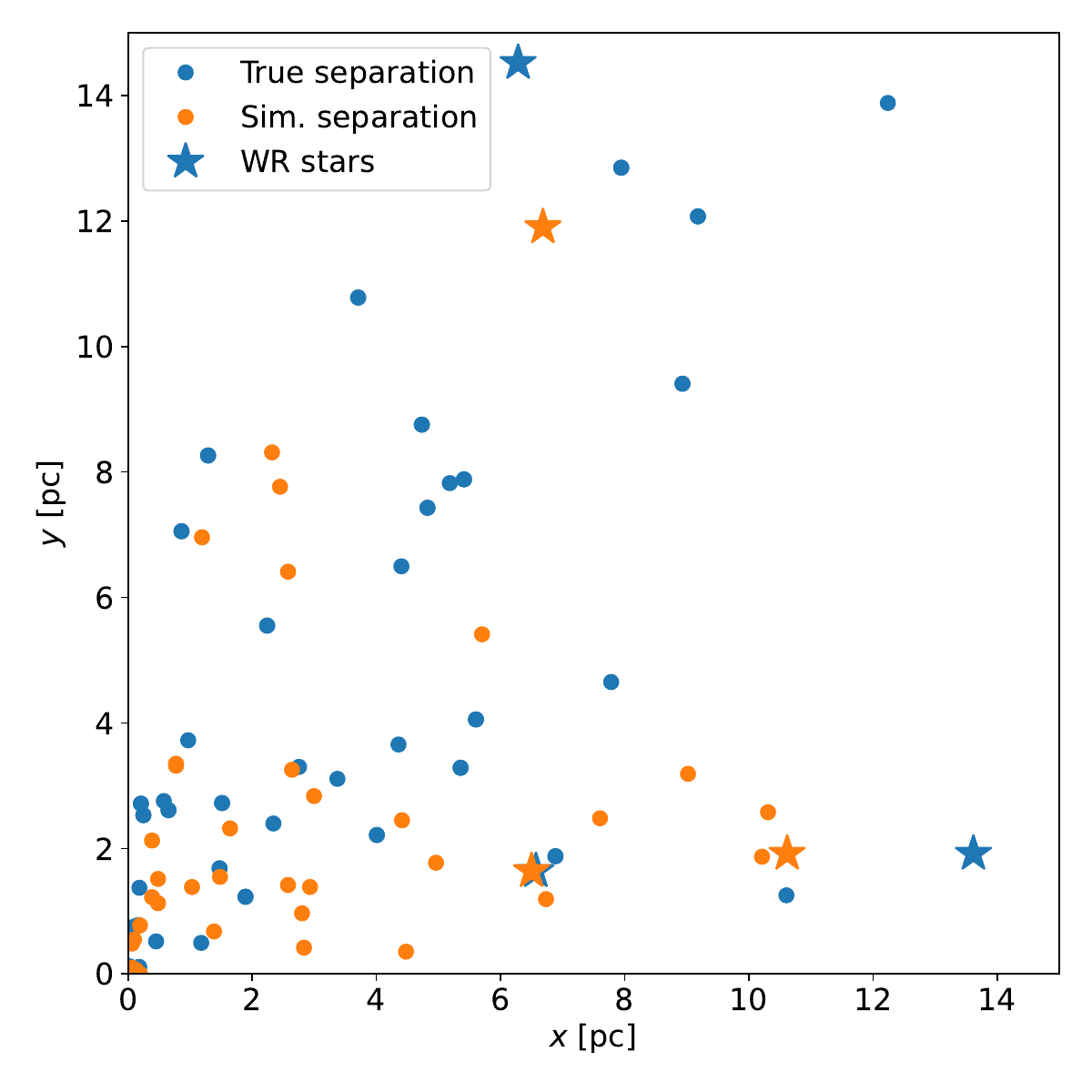}
   \caption{Comparison of the absolute $x$ and $y$ separations for the ``True'' values given in \citet{Berlanas2020} (blue markers, assuming a distance to the cluster of 1.65~kpc) compared with the separations of the 36 O stars in our simulation (orange). We also show the ``True'' positions of the WR stars with star markers in blue, and their shifted positions in the simulation in orange. A 3D visualisation of the simulated cluster is shown in Fig.~\ref{fig:clusterconfig3D}.}
   \label{fig:position_comp}
 \end{figure}

Unfortunately, reliable Gaia parallaxes only exist for just over half of the sample \citep{Berlanas2020}. We therefore need to extrapolate from the right ascension and declination in order to determine a position in 3D.
To produce a list of 3D coordinates for each of our stars, we first define a pair of $(x,y)$ coordinates with respect to the centre of Cygnus OB2, which are calculated using Gaia right ascension and declination assuming a distance of 1.65~kpc to the centre of the cluster. We then extrapolate a 3D distance $d = \sqrt{x^2 + y^2 + \((\lvert x \rvert + \lvert y \rvert)/2\)^2} $. This procedure provides a conservative estimate of the real distance, which could actually be much larger in 3D. Finally, in order to avoid a biased 3D reconstruction, in particular to produce a distribution that is statistically spherically symmetric, we randomly pick a point on the sphere located at the distance $d$ from the cluster centre.
By construction, this reproduces the radial distribution of stars in the cluster, while allowing us to make sensible estimates for the z-component. Fig. \ref{fig:position_comp} shows that the projected distribution of stars in the projected plane resembles qualitatively that given in \citet{Wright2015}.

To accurately resolve massive stars individually in a simulation requires a high resolution. Extending this resolution over the entire domain would be computationally too expensive. Therefore we need to define a ``core region'', at $\vert x \vert , \vert y\vert ,\vert z\vert  < \Delta_c$, such that stars located outside of the cube of half edge-length $\Delta_c$ will be excluded from the simulation. Our computational resources allow us to set $\Delta_c = 12$~pc. As seen in Fig.~\ref{fig:astrometry}, this is enough to include most of the powerful stars, with the notable exceptions of WR145 and WR146, which, with a mechanical power of respectively $2.1 \times 10^{37}$~erg/s and $3.4 \times 10^{37}$~erg/s, are the second and third most powerful stars in Cygnus OB2, thus should not be discarded. We artificially bring them closer, shifting the $y$ coordinate of WR145 from 14.5~pc to 11.9~pc, and the $x$ coordinate of WR146 from 13.6~pc to 10.6~pc. For these two stars, we set $z=0$ without loss of generality.

Stars which are not powerful enough to expand a supersonic wind against the ISM pressure beyond the injection cells cannot be included in the simulation \citep{pittard2021}: we must discard stars such that $\mdot \vinf < (2000~\textrm{km/s}) \times (10^{-7}~M_\odot\textrm{/yr})$. This represents a population of 20 O stars which has a negligible mechanical power ($1.8 \times 10^{36}$~erg/s) and is therefore not expected to affect the cluster feedback at large scales.

Removing all O stars located outside of the cube of half edge-length 12~pc, we end up with a population of 36 O stars, which has a total mechanical power of $7.7 \times 10^{37}$~erg/s, 73\% of which being blown by the most powerful O stars \#7, \#8, \#11, \#22. Cygnus~OB2-8B and Cygnus~OB2-8C are located too close to each other (0.12 pc) to be individually resolved in the inner parsec, so we merge them together by adding up their mass-loss rate and mechanical power.

\begin{figure}
          \centering
              \includegraphics[width=\linewidth]{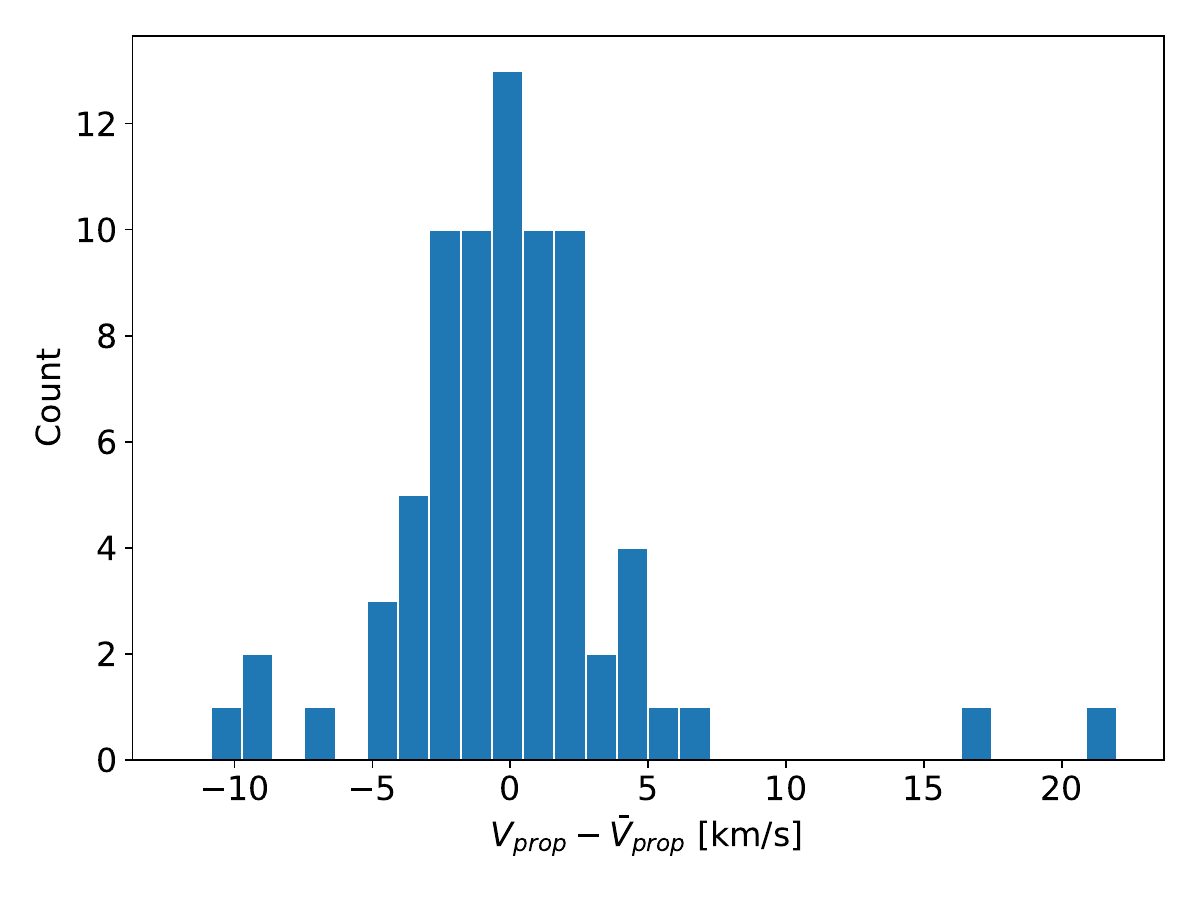}
   \caption{Histogram of individual stars' projected proper motion $V_{\text{prop}}$ minus the mean proper motion of the sample $\Bar{V}_{\text{prop}}$ based on Gaia DR3 \citep{GaiaDR32023}.}
   \label{fig:propermotion}
 \end{figure}

Fig.~\ref{fig:propermotion} shows the velocity dispersion of the stars in the \citet{Berlanas2020} sample. We see that the vast majority of the stars are moving at no more than a few $\kms$, corresponding to a proper motion of order $\sim 4 \,\mathrm{pc}/\mathrm{Myr}$. Since it typically takes a few hundreds of kyrs to reach a quasi-stationary state in the simulation, it is reasonable to neglect the proper motion of individual stars. Besides, as discussed in detail in \citet{wright2016}, observations show no hint of a radial expansion of the cluster: it is unlikely that Cygnus~OB2 was more compact in the past.

Finally we set up an additional O star near the centre of the cluster with fiducial wind parameters $\mdot = 10^{-6} M_\odot$/yr, $\vinf = 2500$~km/s. This star is introduced as a putative supernova progenitor, which will be used in a follow-up work. In the present simulation, it is kept in the main-sequence and has a negligible impact on the results.

The stars forming our simulated cluster are listed in Appendix~\ref{appsimulatedcluster}, and Fig.~\ref{fig:clusterconfig3D} displays the reconstructed 3D distribution. Although this population is not \textit{exactly} that of Cygnus OB2, it is representative of the real statistical distribution of the wind mechanical power in 3D space and, as discussed earlier, was designed to maximise the feedback when the parameters were uncertain. This synthetic population can only be used to draw qualitative conclusions on the gas density and flow configuration across a few tens of parsecs. Our aim is not to perform a quantitative comparison between the simulation results and specific observations such as X-ray data or gas density maps, but rather to probe the presence of a large-scale cluster wind termination shock, which is key to understanding the gamma-ray emission from the region.

\begin{figure*}
          \centering
              \includegraphics[width=0.89\linewidth]{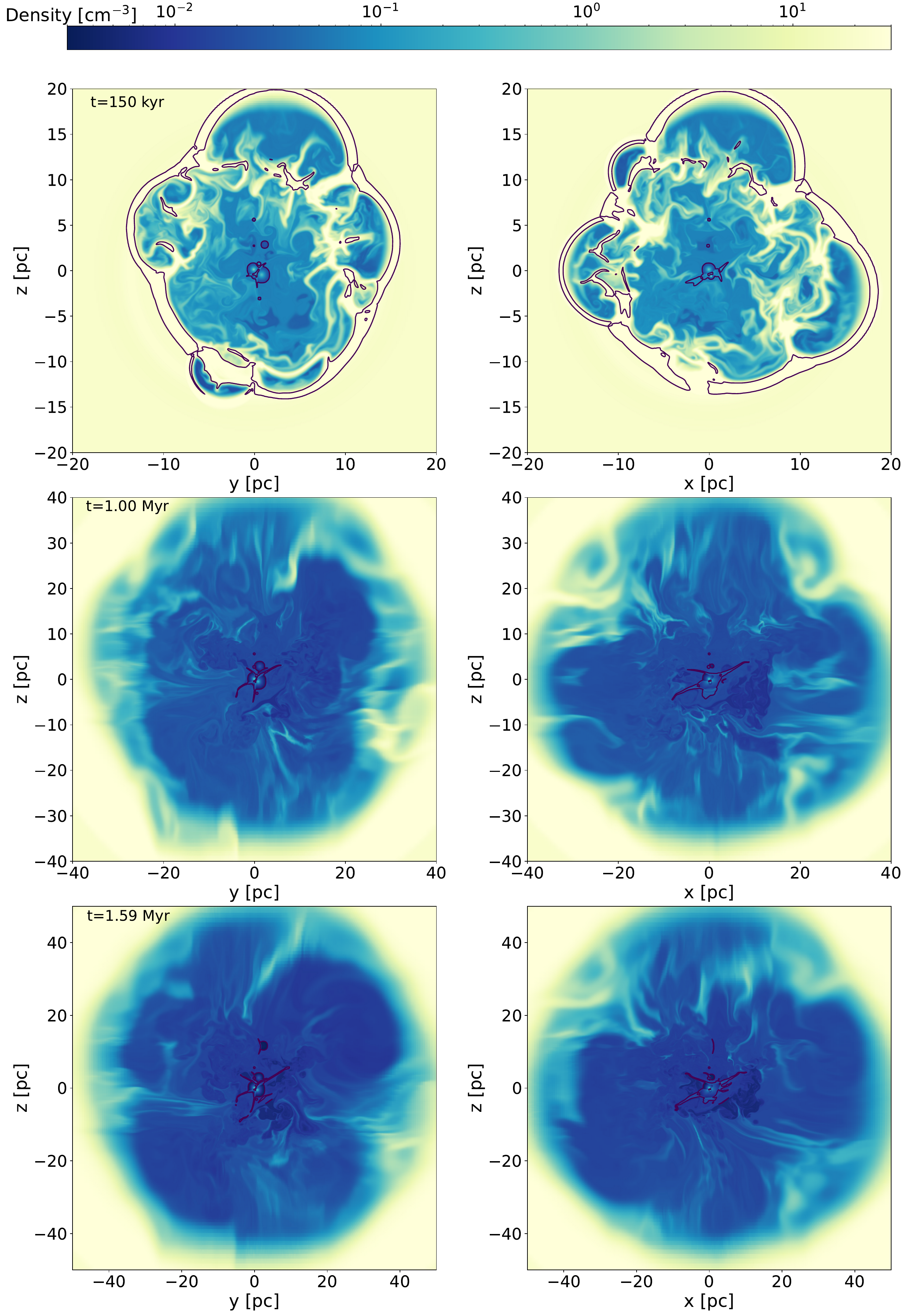}
   \caption{Two different 2D slices at $x=0$ (left) and $y=0$ (right) showing the density map at 3 different times. The purple outlines highlight the Mach=1 contours.}
   \label{fig:timeevolution_density}
 \end{figure*}

\subsection{Initial conditions and stellar wind injection}
The 3D simulation domain extends in the three Cartesian axes from -70~pc to +70~pc. This is large enough to contain the superbubble blown by the cluster during about 2~Myr. We set up 776$^3$ cells over the core region defined by -12.5~pc~$< x, y, z < 12.5$~pc, therefore using a resolution of 0.032~pc/cell in this region. The grid is stretched beyond 12.5~pc up to the boundaries of the domain, degrading progressively the resolution, eventually using a total number of 1000$^3$ cells.

The initial condition is an ideal background gas of number density 20~cm$^{-3}$, mean molecular weight 0.61 and temperature $10^4$~K. These values are compatible with recent observational analyses of the molecular clouds in the region \citep[e.g.][]{zhang2024,Astiasarain2023,lhaasocygnus2024}, with typical HI column density measurements on the order of $10^{22}~$cm$^{-2}$. Given the complex structure of the region, it is difficult to accurately isolate the contribution of the molecular gas near Cygnus~OB2. We choose a density typical of a parent molecular cloud which is expected to form a massive star cluster.

Setting a somewhat high external density is expected to inhibit the expansion of the superbubble blown by the massive stars, which is necessary to maintain the entire structure inside the numerical box over the simulation time (2~Myr). On the other hand, one should ensure that the pressure inside the superbubble remains larger than the external pressure over the simulation time, which is always true with our assumed parameters (at $t=2$~Myr the internal pressure is still more than twice the external pressure).

Within the high-resolution core, each massive star is individually modelled as a spherically symmetric stellar wind of terminal velocity $\varv_{\infty,i}$ and mass-loss rate $\dot{M}_i$:
\begin{align}\label{injectionprofile}
    \v{u} = \varv_{\infty,i} \v{e_{r,i}} \, , \qquad 
    \rho = \frac{\dot{M}_i}{4 \pi r^2 \varv_{\infty,i} } \, ,
\end{align}
where the parameters $\varv_{\infty,i}$ and $\dot{M}_i$ are summarised for each star in Table~\ref{tab:listsimulatedappendix} and are kept constant in time. The pressure in the wind is prescribed assuming an isothermal wind: $p_{\rm gas} = \rho c_s^2$, with $c_s = 20$~km/s for O stars and $c_s = 30$~km/s for WR stars. During the simulation, the stellar winds are set up within spheres of radius $R_H = 0.16$~pc (5 cells) around each star, which is almost five times smaller than the distance between the two closest stars, allowing to properly resolve the regions of stellar wind collisions (except that between Cygnus~OB2~\#8B and Cygnus~OB2~\#8C which have been merged). This also ensures that for all simulated stars, the injection radius is significantly smaller than $r_{\rm inj,max} = \( \mdot \vinf / (4 \upi P_0) \)^{1/2}$, with $P_0$ the initial pressure. This condition is necessary to accurately expand wind-blown bubbles \citep{pittard2021}. For the stars with the lowest $\mdot \vinf$ product, we get $R_H/r_{\rm inj,max}$ = 0.3, while for the most powerful stars, which dominate the stellar feedback at large scales, we have $R_H/r_{\rm inj,max} < 0.1$.

The initial ($t=0$) injection setup is slightly different. The stellar winds are defined over 0.39~pc (12 cells), which corresponds to half the distance between the two closest stars. In this extended injection region, we can define a smoother cut-off for the velocity and the density: $\v{u} = \varv_{\infty,i}/\left(1+(r-R_H)/R_H\right)^2 \v{e_{r,i}}$, $\rho = \mdot_i/ (4 \upi r^2 \Vert \v{u} \Vert ) $. This was deemed necessary in order to obtain spherical stellar winds with the Cartesian grid.

WR stars are only expected to appear at the end of the main-sequence phase. WR progenitors are assumed to be O stars with fiducial wind parameters $\mdot = 10^{-6} M_\odot$/yr and $ \vinf = 2500$~km/s. Since the WR stars are considerably off-centred, their feedback in the main-sequence is not expected to play a major role. In the simulation, the transition to the WR phase happens at $t=1.6$~Myr. We inject the WR winds over 6 cells using Eq.~\ref{injectionprofile}.

The Euler equations of hydrodynamics are solved using the publicly available code PLUTO \citep{PLUTO2007} with a Lax-Friedrichs scheme (TVDLF). Thermal conduction and radiative cooling are neglected in order to minimise the computational overhead in the high-resolution core. Besides, a proper treatment of thermal conduction cannot be done without accounting for the magnetic field. Thermal conduction and cooling are known to impact on the properties and stability of the superbubble shell, to increase the density inside the cavity due to the evaporation at the shell interface and to lower the temperature inside the cavity \citep{weaver1977,elbadry2019}. On the one hand, not encapsulating these effects implies that our simulation is not expected to provide reliable results for the dynamics of the superbubble shell, nor to provide quantitative predictions for the density and temperature, which could be confronted to radio, optical or X-ray observations. On the other hand, neither thermal conduction nor cooling are expected to substantially change the pressure inside the superbubble, which is the main driver for the formation of the cluster WTS. For the purpose of this work, it is therefore reasonable to neglect these processes.

\section{Results}
\label{sec:results}
\subsection{Structure of the superbubble}
\begin{figure}
          \centering
              \includegraphics[width=\linewidth]{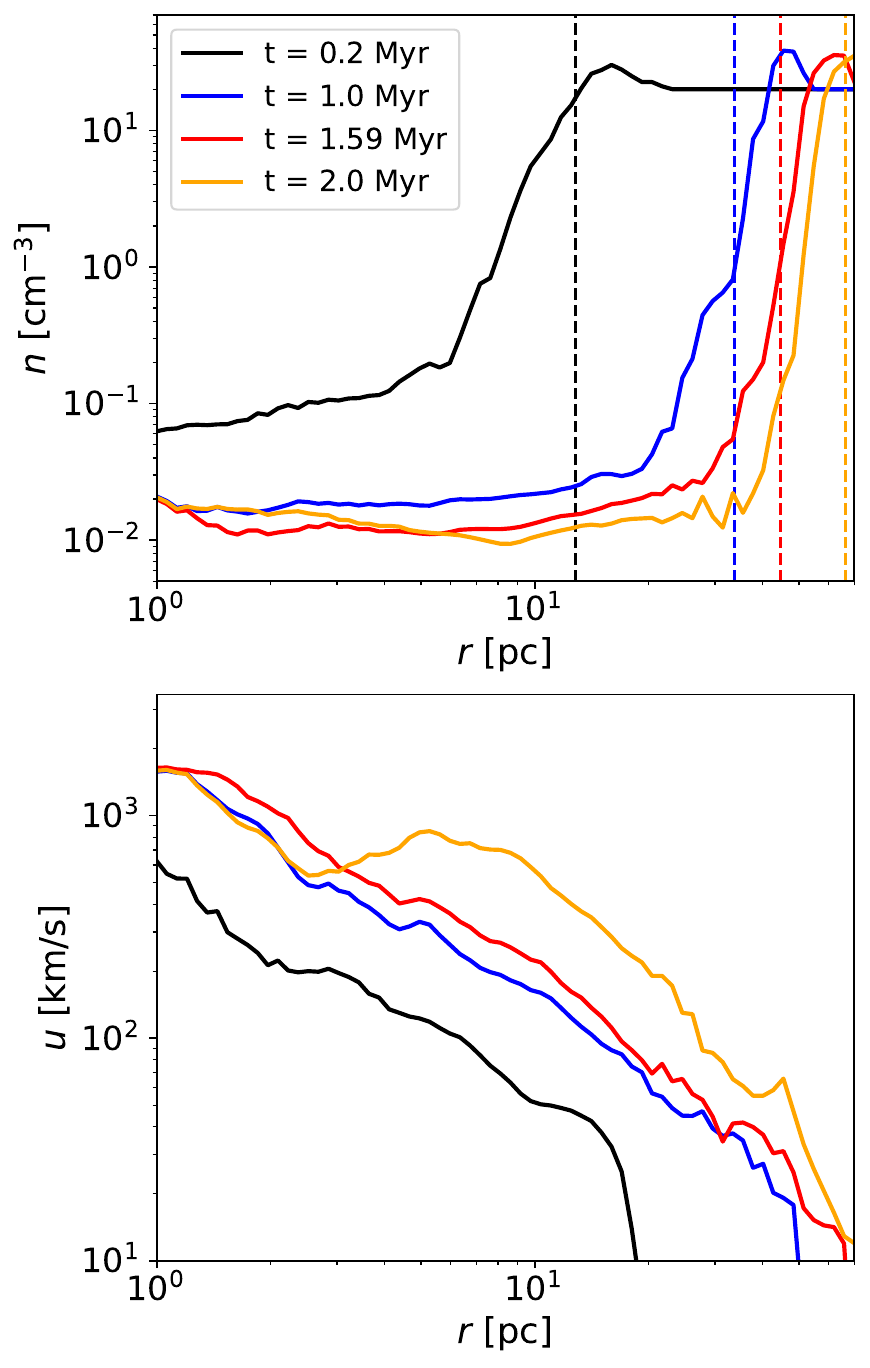}
   \caption{Radial profiles of number density (top panel) and velocity (bottom panel) at 0.2~Myr (early phase), 1~Myr (quasi-stationary state), 1.59~Myr (just before the onset of WR stars) and 2~Myr (end of the simulation). The vertical lines in the top panel show the theoretical position of the superbubble shell according to \citet{weaver1977}.}
   \label{fig:radialprofiles_rhou}
 \end{figure}

Snapshots, using 2D slices through the simulation at 150~kyr (early stage), 1~Myr and 1.59~Myr (just before the onset of WR stars) are shown in Fig.~\ref{fig:timeevolution_density} while averaged radial profiles of density and velocity are plotted in Fig.~\ref{fig:radialprofiles_rhou}. At the beginning of the simulation, each massive star is surrounded by its own wind-blown bubble. The latter expand over several tens of kyrs until they start to percolate and eventually merge to form an approximately spherical superbubble. The matter excavated out accumulates in the superbubble shell, whose position matches the theoretical expectation (see the top panel of Fig.~\ref{fig:radialprofiles_rhou}), although the shell width is much broader than theoretically expected due to the absence of radiative cooling in the simulation. The density inside the superbubble is low, of the order of 0.01~cm$^{-3}$, although accounting for thermal conduction would trigger mass loading from the shell and increase this value.

\begin{figure*}
          \centering
              \includegraphics[width=\linewidth]{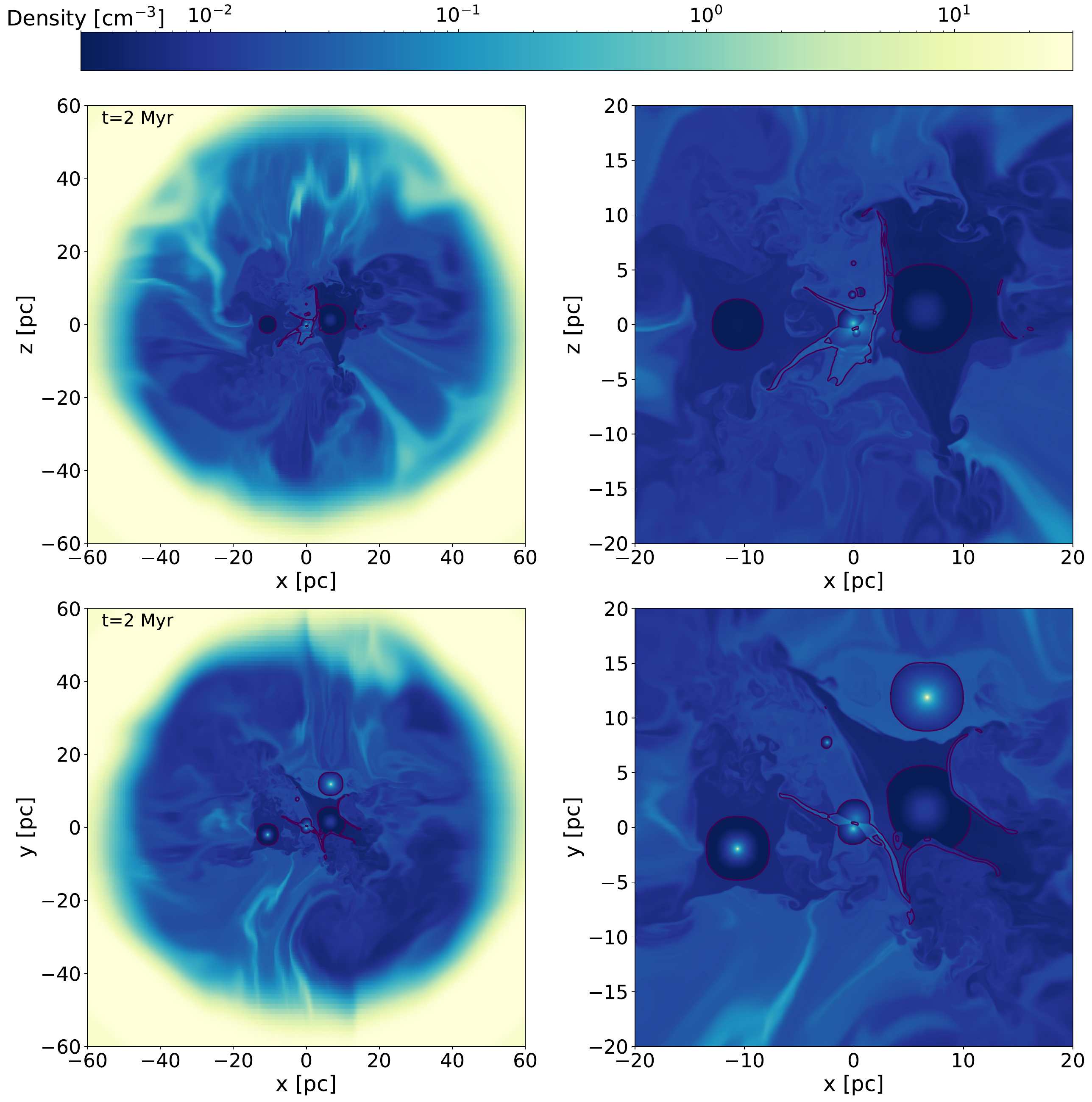}
   \caption{Density maps at 2~Myr (400~kyr after the onset of WR stars) in two different slices at $y=0$ (top) and $z=0$ (bottom). Left panels: full view of the superbubble. Right panels: zoomed-in views. The red outlines show the Mach=1 contours.}
   \label{fig:maps_density_at_2Myr}
 \end{figure*}

From Fig.~\ref{fig:radialprofiles_rhou} we see that the radial profiles averaged over the azimuthal angles are relatively smooth, in contrast with theoretical expectation \citep[see e.g. Fig.~1 of][]{gupta2018b}. This is understood since we do not have a spherical wind around a compact central cluster. The winds blown by the powerful O stars in the inner few parsecs (namely Cygnus~OB2~\#7, Cygnus~OB2~\#8B+\#8C and Cygnus~OB2~\#22) are strongly asymmetric. The supersonic outflows are shaped, through wind-wind collisions, into two-dimensional sheets, and then possibly, via subsequent collisions with other stellar winds, to one-dimensional jets that extend over large distances (up to 10~pc).

Overall, the level of wind-wind interaction is very low. At this stage, most of the stars are still surrounded by a small-scale ($\approx 1$~pc) stellar wind termination shock, which slowly expands in the low-density cavity. 

After 1~Myr, the simulation has already reached a quasi-stationary state, in the sense that we do not witness qualitative changes anymore. Due to the high external pressure, the forward shock of the superbubble has already become subsonic. The expansion of the superbubble follows the expected $t^{3/5}$ scaling \citep{weaver1977}. Individual wind termination shocks, trans-sonic sheets and jets continue to expand slowly, in an almost self-similar manner.

Wolf-Rayet stars are introduced at 1.6~Myr and blow powerful winds until the end of the simulation. Fig.~\ref{fig:maps_density_at_2Myr} shows the final snapshot at 2~Myr. At this point, the Wolf-Rayet stars are expected to either explode in supernovae or collapse into black-holes. The presence of WR stars has overall little impact on the dynamics of the superbubble. This is not surprinsing since the size of the superbubble shell theoretically scales as $L_c^{1/5}$, where $L_c$ is the total wind power of the cluster. Since the three WR stars are strongly off-centred, their feedback actually acts \textit{against} the expansion of the wind termination shocks from the central region. Indeed, their onset provokes the shrinking of the individual O star WTSs. Finally, 400~kyr after they have appeared, they are still far away from interacting despite having blown their own individual WTSs over a couple of parsecs.

\begin{figure}
          \centering
              \includegraphics[width=\linewidth]{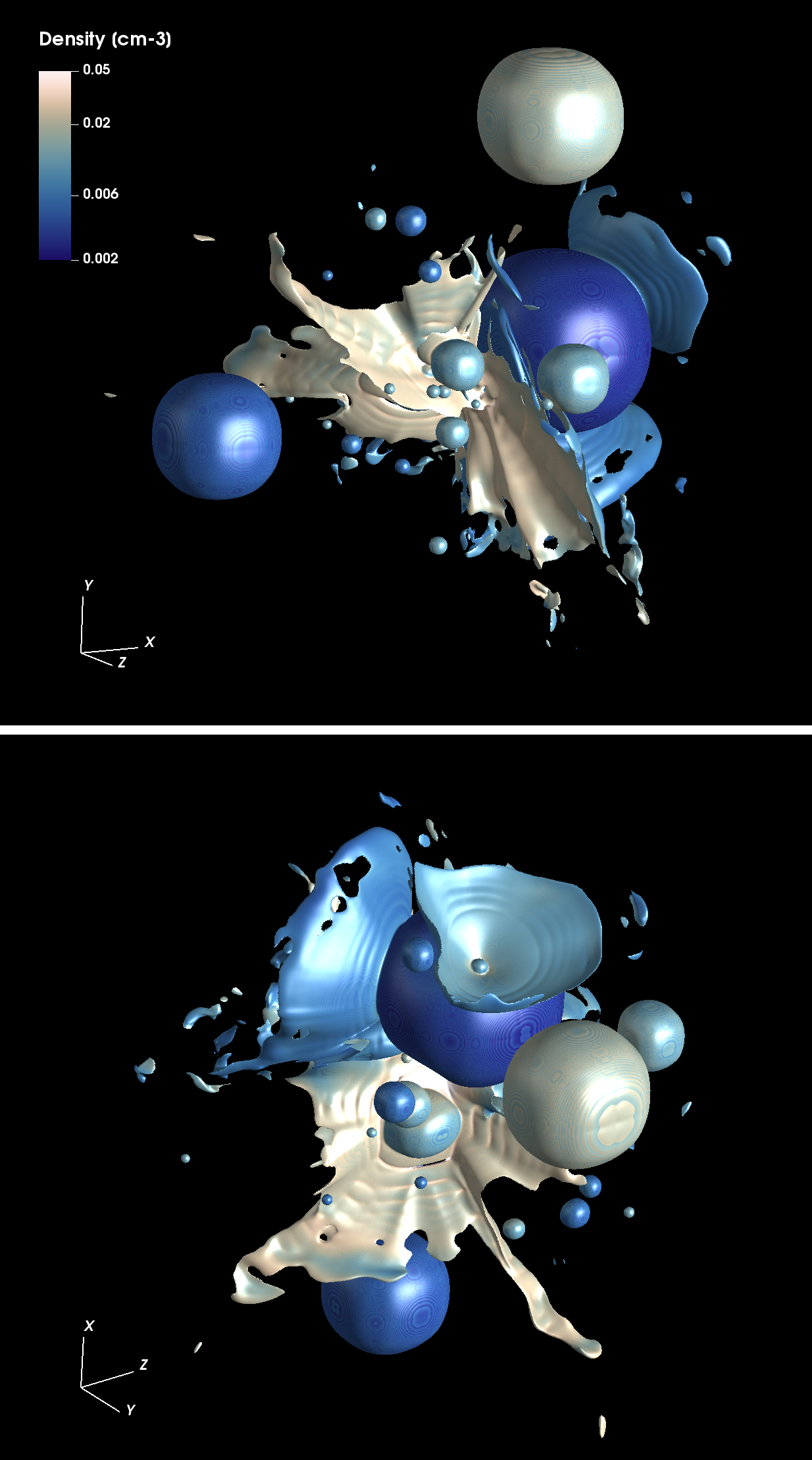}
   \caption{3D views of the cluster core at 2~Myr, showing the isosurfaces at Mach=1. The colorscale shows the density on these surfaces. See also the corresponding movie in the online Supplementary Material.}
   \label{fig:Cygnus3Dviews}
 \end{figure}

Fig.~\ref{fig:Cygnus3Dviews} provides a 3D view of the cluster at 2~Myr showing the isosurfaces of Mach number equal to 1 (see also the corresponding movie in the online Supplementary Material). The spheres correspond to individual stellar WTS around single stars. It is clearly seen that only a handful of powerful stars dominate the mechanical feedback. Most O stars are not powerful enough to expand a stellar WTS beyond one parsec: there are seen as isolated small spheres on the 3D views. These views provide a better understanding of the flow configuration and highlight in particular the absence of any large-scale spherical shock in the system, apart from individual WR shocks which extend over a couple of parsecs, as theoretically expected. Wind-wind collisions are clearly seen in particular between WR stars and O stars. These collisions produce trans-sonic planes, possibly curved inward the less powerful star. Further interaction with multiple stars sculpt the trans-sonic flows. We witness in particular the formation of a jet in the $(x,y)$ plane, as can also be seen in lower panels of Fig.~\ref{fig:maps_density_at_2Myr}.

\subsection{Understanding the absence of a large-scale cluster wind termination shock}
It is usually assumed that if the putative cluster WTS is larger than the cluster core, then it will actually form beyond the boundary of the core \citep[e.g.][]{vieu2023,menchiari2024}. It appears that this argument is too naive. Firstly, a high number of massive stars organised in several layers is required to trap the kinetic energy of the winds within a rather small region of energy deposition, and eventually produce a collective, radial, outflow. If the number of stars is too small, the hydrodynamics is dominated by a few wind-wind collisions which create highly asymmetric structures. Secondly, the stars must be close enough to each other in order for their winds to interact. Indeed, the kinetic energy drops rapidly (typically as $r^{-4}$) beyond the WTS of a single star, which implies that the kinetic energy injected by a single star is contained within its own WTS. The latter expands in the shocked plasma which is nearly statistically homogeneous inside the superbubble. In order to enable efficient wind-wind interactions, the individual wind termination shocks must be close enough to one another, something which is not the case if the average distance between the stars greatly exceeds the typical size of a WTS blown by an O star. The textbook scenario of a continuous extended deposition of energy \citep{chevalier1985} actually describes the extreme case where the wind-wind interactions are so efficient that all the stellar wind kinetic energy is converted to thermal pressure within the core. Although this might apply in some extremely compact and powerful clusters such as Westerlund 1 \citep[e.g.][]{haerer2023}, it is not the case for Cygnus~OB2.

The average pressure inside the superbubble (in the shocked plasma downstream of individual wind termination shocks) scales as follows \citep{weaver1977}:
\begin{equation}\label{pressureweaver}
    P_{SB} \approx 0.16 L_c^{2/5} \rho_0^{3/5} t^{-4/5} \, ,
\end{equation}
where $L_c$ is the total power of the cluster, $\rho_0$ is the initial density and $t$ the age of the cluster. This relation is derived by imposing energy and momentum conservation in the entire superbubble and should therefore hold independently of the spatial distribution of the stars inside the superbubble.
In order to estimate the position of the stellar WTS of a single star isolated in the superbubble, one has to balance the ram pressure of the stellar wind, $P_w = \mdot \vinf/(4 \upi R_\text{wts}^2)$ with the pressure inside the superbubble. This provides:
\begin{multline}\label{singlestarwtssize}
    R_\text{wts} \approx 1.5 \left(\frac{\mdot}{ 10^{-6} M_\odot \rm /yr  }\right)^{1/2} \left( \frac{\vinf}{2500~\rm km/s}\right)^{1/2} \\
    \times \left(\frac{L_c}{2 \times 10^{38} {\rm ~ erg/s}}\right)^{-1/5} \left(\frac{t}{5~\rm Myr}\right)^{2/5}
     \left(\frac{\rho_0}{20 m_p~{\rm cm}^{-3}}\right)^{-3/10}  \rm pc
    \, ,
\end{multline}
where $\mdot$ is the mass-loss rate of the isolated star and $\vinf$ its wind velocity. We have normalised the parameters to values typical of Cygnus~OB2. In fact, $R_\text{wts}$ depends only weakly on the parameters, which means it is quite general to state that a typical O star isolated in \textit{any} superbubble will blow a supersonic wind over at most a few parsecs. In order for stellar winds to interact, the typical distance between the O stars must therefore be less than a few parsecs.

\begin{figure}
          \centering
              \includegraphics[width=\linewidth]{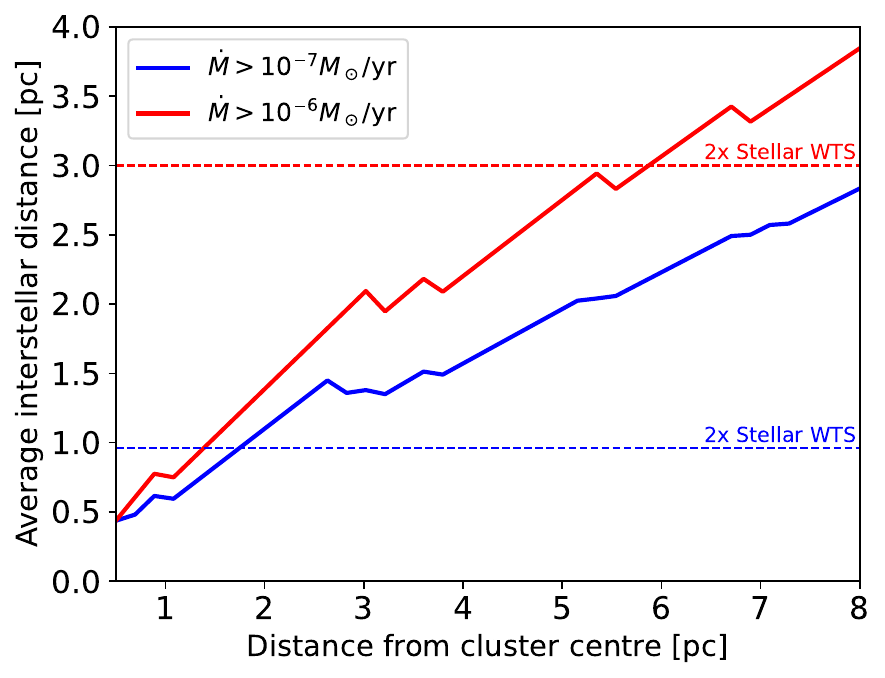}
   \caption{Average distance between the stars within a given radius for stars with $\mdot > 10^{-7} M_\odot$/yr (nearly all simulated stars) and stars with $\mdot > 10^{-6} M_\odot$/yr (8 stars). Wolf-Rayet stars are not included. The red dashed line shows the theoretical size of the WTS blown by a 5~Myr old star with $\mdot = 10^{-6} M_\odot$/yr and $\vinf = 2500$~km/s. The blue dashed line shows the theoretical size of the WTS blown by a 5~Myr old star with $\mdot = 10^{-7} M_\odot$/yr and $\vinf = 2500$~km/s.}
   \label{fig:WTScriterion}
 \end{figure}

In Cygnus OB2, the distribution of stars is not homogeneous but peaked around the centre, as can be seen in the middle panel of Fig.~\ref{fig:astrometry}. Therefore, the average distance between O stars increases as one moves away from the cluster centre. Fig.~\ref{fig:WTScriterion} shows the evolution of the average distance between the stars, $d = (2/3)^{1/3} N(r)^{-1/3} r$, where $N(r)$ is the number of O stars within the radius $r$ in our simulated sample. This should be representative of Cygnus OB2. Efficient wind-wind interactions are expected only if a single star WTS is a sizeable fraction of the average distance between the stars. One concludes that, in our sample, only the most powerful O stars ($\mdot > 10^{-6} M_\odot/\mathrm{yr}$) within a distance of a couple of parsecs from the cluster centre, can interact efficiently. The other, less powerful, stars can only interact within the inner few parsecs. This explains why wind-wind interactions are only efficient in the inner few parsecs in the simulation and dominated by the most powerful stars.

The argument exposed above is quite general and the conclusion  depends only weakly on the parameters of the problem. In particular, even if the real configuration of the stellar population in Cygnus OB2 is slightly different than the one we simulate, or if it happens that the population was different in the past, or if the initial density is lower than the value assumed in the simulation, one would still not expect to see a large-scale WTS in the region.

\begin{figure}
          \centering
              \includegraphics[width=0.9\linewidth]{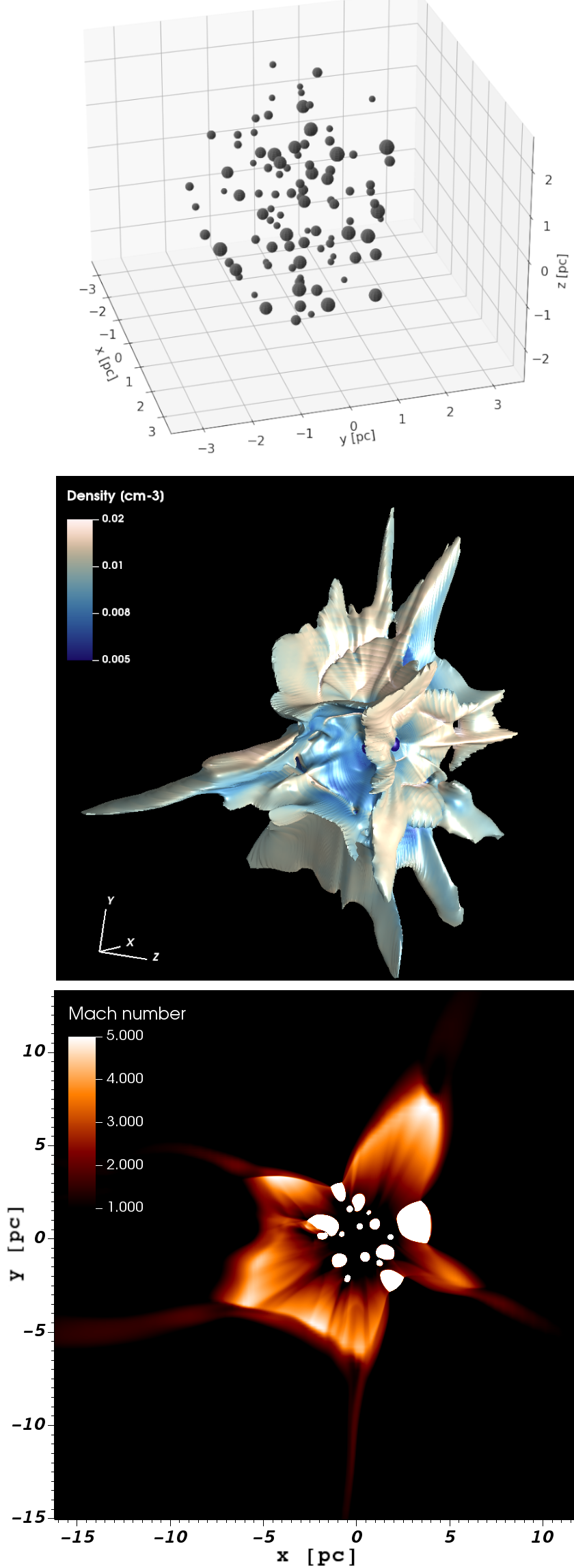}
   \caption{Result of a complementary simulation with 100 massive stars, from $24 M_\odot$ to $90 M_\odot$, sampled from a Salpeter IMF and uniformly gathered within 3~pc. Top panel: 3D distribution of the stars. The size of the spheres scales as $L_w^{1/5}$, with $L_w$ the mechanical power of the star. Middle panel: Mach=1 isosurfaces at 2~Myr. Bottom panel: Mach number in the $(x,y)$ plane ($z=0$) at 2~Myr.}
   \label{fig:extendedsimu}
 \end{figure}

As rule of thumb, a massive star cluster could create a large-scale WTS only if the cluster core is smaller than a few parsecs. However, this might not be a sufficient condition to generate a \textit{spherical} wind termination shock. In order to investigate this last point, we ran a fiducial simulation of a more compact cluster, with a synthetic population of 100 massive stars generated from a Salpeter IMF and estimating the wind parameters using Eqs.~\ref{stellarparametersMdot},\ref{stellarparametersVw}. The spatial distribution is randomly generated assuming a uniform distribution of stars within 3~pc, in an ISM of initial density 10~cm$^{-3}$. The total power of this fiducial cluster is $1.1 \times 10^{38}$~erg/s. If such a power would be injected as a point-like source, one would expect a spherical cluster WTS to develop up to 13~pc after 2~Myr. Fig.~\ref{fig:extendedsimu} shows the simulation result after 2~Myr. Clearly, wind-wind interactions are much more efficient in such a compact case than it is in the case of Cygnus OB2. These interactions create a number of trans-sonic sheets and jets, which are still able to escape from the core region. Therefore, even in such a compact case, the wind kinetic energy escapes the region of energy deposition in a very inhomogeneous manner and the shock surface shape is irregular. This highlights the difficulty to generate a spherical cluster wind termination shock from a population of O stars that is only \textit{statistically} spherically symmetric. 

A detailed study of the structure of the shocks and jets in such a compact cluster is left for future work.

\begin{figure}
          \centering
              \includegraphics[width=\linewidth]{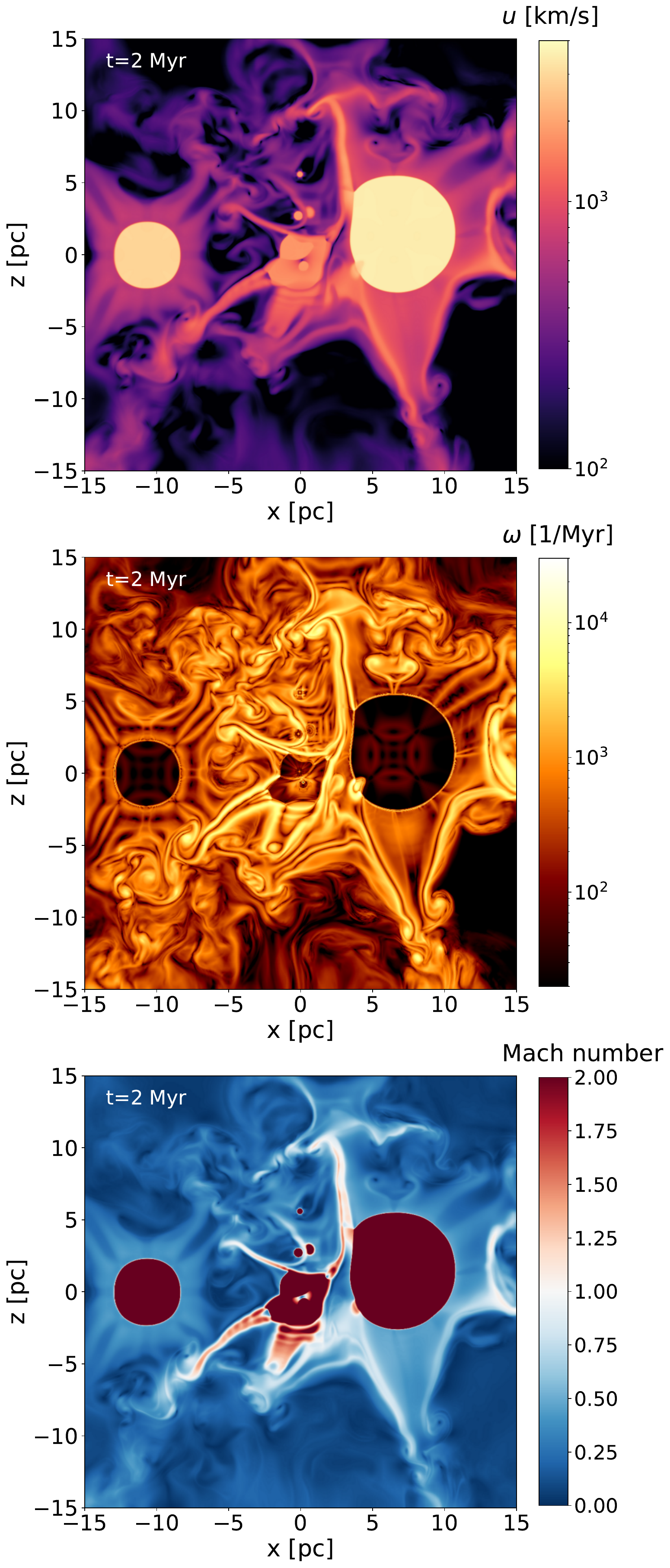}
   \caption{Wind-wind interactions in the inner region at 2~Myr. Top panel: magnitude of the velocity field. Middle panel: magnitude of the vorticity. Bottom panel: Mach number. Dark red shows regions of Mach number larger than 2. The largest WTS on the left and right are respectively blown by WR146 and WR144. The slices are taken at $y=0$.}
   \label{fig:innerregion}
 \end{figure}

\subsection{Consequences for high-energy astrophysics}
The aim of the present study is to diagnose the shocks in the Cygnus OB2 star cluster, which are key to understanding the high-energy emission observed from the region. Our simulation shows that a scenario based on a large-scale spherical cluster WTS is excluded. On the other hand, the  powerful Wolf-Rayet stars do blow strong large-scale spherical shocks which could in principle be sites of particle acceleration \citep[e.g.,][]{casse1980,ptuskin2018}. In the simulation, we ran the Wolf-Rayet phase during 400~kyr, which corresponds to the typical lifetime of central Helium burning in a massive star \citep[e.g.]{Woosley2019,Higgins2021}. Given that some stars might only reach the WR stage later in their central He burning \citep[e.g.,][]{Josiek2024}, this can be considered a upper limit for the typical WR lifetime. At this point, the most powerful WR star has expanded a shock up to a radius of 5~pc.

We can work out an upper bound on the maximum achievable energy of a proton accelerated by the WTS around WR144. Efficient particle acceleration requires a super-Alfvénic shock, which limits the upstream magnetic field as $B R_\text{wts} < \sqrt{\vinf \mdot} $. Besides, the Hillas criterion \citep{hillas1984} sets an upper bound on the maximum energy as $E_\text{max} < q B R_\text{wts} \vinf/c$, which is actually the limit imposed by adiabatic losses in the radial wind and therefore cannot be overcome by enhanced magnetic field amplification downstream of the shock e.g. due to the Cranfill effect or nontrivial interactions with the nearby O star winds. We eventually get $E_\text{max} < 1.5 Z$~PeV for the wind parameters of WR144 ($\mdot = 2.4 \times 10^{-5} M_\odot$/yr, $\vinf = 3500$~km/s). This should be taken as a strict upper bound and not an estimate, for we know that the limiting factors on the maximum achievable non-thermal energy around a spherical shock are more stringent than the Hillas limit \citep{morlino2021}, and the efficiency of particle acceleration at quasi-perpendicular shocks is debated \citep{caprioli2014,xu2020,kumar2021}. On the other hand, a regime of perpendicular diffusion around the WTS might enhance the maximum achievable energy \citep{Jokipii1987}, although this requires a detailed treatment of the transport along the shock surface in a Parker-spiral magnetic field \citep{Kamijima2022}. In the end, even if the WR stars in Cygnus OB2 are unlikely to account alone for the ultra-high energy gamma-rays detected from the Cygnus region, they might provide a non-negligible flux of high-energy particles at the level of a fraction of the wind power up to hundreds of TeV.

In addition to the individual stellar wind termination shocks, there is still some level of collective interaction in the inner few parsecs, mostly due to the stars WR144, Cyg~OB2~\#7, Cyg~OB2~\#8B+\#8C, Cyg~OB2~\#22 and a couple of less powerful O stars. As can be seen in Fig.~\ref{fig:innerregion}, these stars are close enough to interact efficiently, creating structured trans-sonic sheets and jets. The shocked plasma is characterised by a high velocity (hundreds of km/s) and high vorticity, which suggests a high level of hydrodynamic turbulence with the formation of large-scale eddies. Particle acceleration in a collection of strong shocks and strong turbulence in Cygnus~OB2 was investigated by \citet{bykov2022}. The authors claimed that this model was able to reproduce the observations up to the highest energies, assuming a cluster size of 55~pc, a turbulent velocity of 1500~km/s in the entire region. Indeed, these parameters together with the assumption of a 15~\textmu G magnetic field provide a maximum energy of about 4~PeV, assuming that the Hillas limit applies, which is not unlikely in this acceleration scenario \citep{vieu2022Emax}. However, Fig.~\ref{fig:innerregion} shows that this choice of parameters is not realistic. The turbulent region extends at most over 10~pc and the mean velocity is only a few hundreds of km/s, which would provide a maximum energy of at most 100~TeV. Furthermore, the purely stochastic model of \citet{bykov2001} does not properly describe the plasma in star cluster environments, because the flow remains laminar inside the stellar wind cavities, which inhibits re-acceleration effects \citep{vieu2024core}. Finally, we stress again that direct wind-wind collisions are not expected to enhance the maximum energy achieved by accelerated particles when a realistic flow configuration in multi-dimensions is taken into account \citep{vieu2020,vieu2024core}.

\section{Conclusions}\label{sec:conclusion}
We analysed the stellar population of Cygnus~OB2 as provided by observations, highlighting that i) most of the cluster power is actually provided by 3 WR stars and 5 O stars, and ii) the distribution of the massive stars extends up to more than 10~pc without a strong mass segregation. Cygnus~OB2 must therefore be considered as a ``loose association'' rather than a ``compact cluster''. In this case, one-dimensional modelling of the cluster feedback is not expected to provide reliable results. In particular, assuming a continuous deposition of thermal energy within a spherical region is an oversimplification. Detailed hydrodynamic simulations are then necessary to properly understand the stellar feedback on interstellar scales.

To this aim, we simulated the 40 most powerful stars in Cygnus~OB2. We were able to resolve each massive star individually in the core, while allowing their feedback to develop features up to several tens of parsecs. Since the uncertainties on the measurements of parallaxes and stellar wind parameters prevent us from modelling the exact stellar population, we set up a cluster whose statistical properties match that of Cygnus~OB2. Our 3D reconstruction aims to reproduce the radial distribution of massive stars, although it is expected to provide a slightly more compact configuration.

Despite these shortcomings, the simulation results remain reliable on the qualitative level. In particular, the absence of a large-scale cluster WTS in the simulation implies the absence of such a cluster WTS in Cygnus~OB2. This result is a natural consequence of the loose distribution of the most powerful stars in space. Since the average interstellar separation is typically much larger than the size of an individual stellar WTS, there is a relatively low level of wind-wind interactions. The kinetic energy of the winds is found to thermalise in a highly inhomogeneous manner which cannot be accounted for using standard spherically symmetric models.

In order to further probe the formation of a cluster WTS, we ran a complementary simulation of a synthetic population of 100 massive stars, clustered in a core with a radius 4 times smaller than that of Cygnus~OB2. We found that even in this case, non-trivial wind-wind interactions on sub-parsec scales prevent the formation of a spherical WTS. Wind-wind collisions rather produce trans-sonic sheets and jets which disperse the kinetic energy to large distances, hindering the expansion of a collective radial outflow beyond the cluster core.

The physical reason underlying the absence of a cluster WTS in Cygnus~OB2 is thus well understood and easily generalisable. In principle it would equally apply in a more refined simulation, e.g. including magnetic fields, thermal conduction and cooling, or for a different cluster, provided the distribution of stars in space is similar. This is the only relevant parameter of the problem. The stars in Cygnus~OB2 are -- and most likely always were -- too dispersed to enable strong collective effects.

Some level of collective interactions is nevertheless found in the inner few parsecs of our simulation, in particular after the onset of WR stars. Similarly to the case of the more compact cluster, collisions between a few individual stellar WTS produce focused trans-sonic outflows. This non-trivial feedback creates an intricate environment with multiple shocks. Although these shocks do provide channels for particle acceleration, they are not expected to contribute beyond energies of a few hundred TeV and are therefore unlikely to be the origin of the PeV gamma-rays detected from the Cygnus region by the LHAASO observatory \citep{lhaasocygnus2024}.

\section*{Acknowledgements}
We thank the referee and the editor for useful suggestions.
This work made use of the MHD code PLUTO. The simulation was performed on the HPC system Raven at the Max Planck Computing and Data Facility. We acknowledge J.S. Wang for helpful discussions on the implementation of the simulation. This work has made use of data from the European Space Agency (ESA) mission {\it Gaia} (\url{https://www.cosmos.esa.int/gaia}), processed by the {\it Gaia} Data Processing and Analysis Consortium (DPAC, \url{https://www.cosmos.esa.int/web/gaia/dpac/consortium}). Funding for the DPAC has been provided by national institutions, in particular the institutions participating in the {\it Gaia} Multilateral Agreement. CJKL gratefully acknowledges support from the International Max Planck Research School for Astronomy and Cosmic Physics at the University of Heidelberg in the form of an IMPRS PhD fellowship. CJKL, AS, and VR gratefully acknowledge support by the German Deut\-sche For\-schungs\-ge\-mein\-schaft, DFG\/ in the form of an Emmy Noether Research Group -- Project-ID 445674056 (SA4064/1-1, PI Sander) and the Federal Ministry of Education and Research (BMBF) and the Baden-W\"{u}rttemberg Ministry of Science as part of the Excellence Strategy of the German Federal and State Governments.

\section*{Data Availability}
The initial conditions and output of the simulation may be shared on reasonable request to the corresponding author.

\appendix
\section{List of simulated stars}\label{appsimulatedcluster}
\begin{figure}
          \centering
              \includegraphics[width=\linewidth]{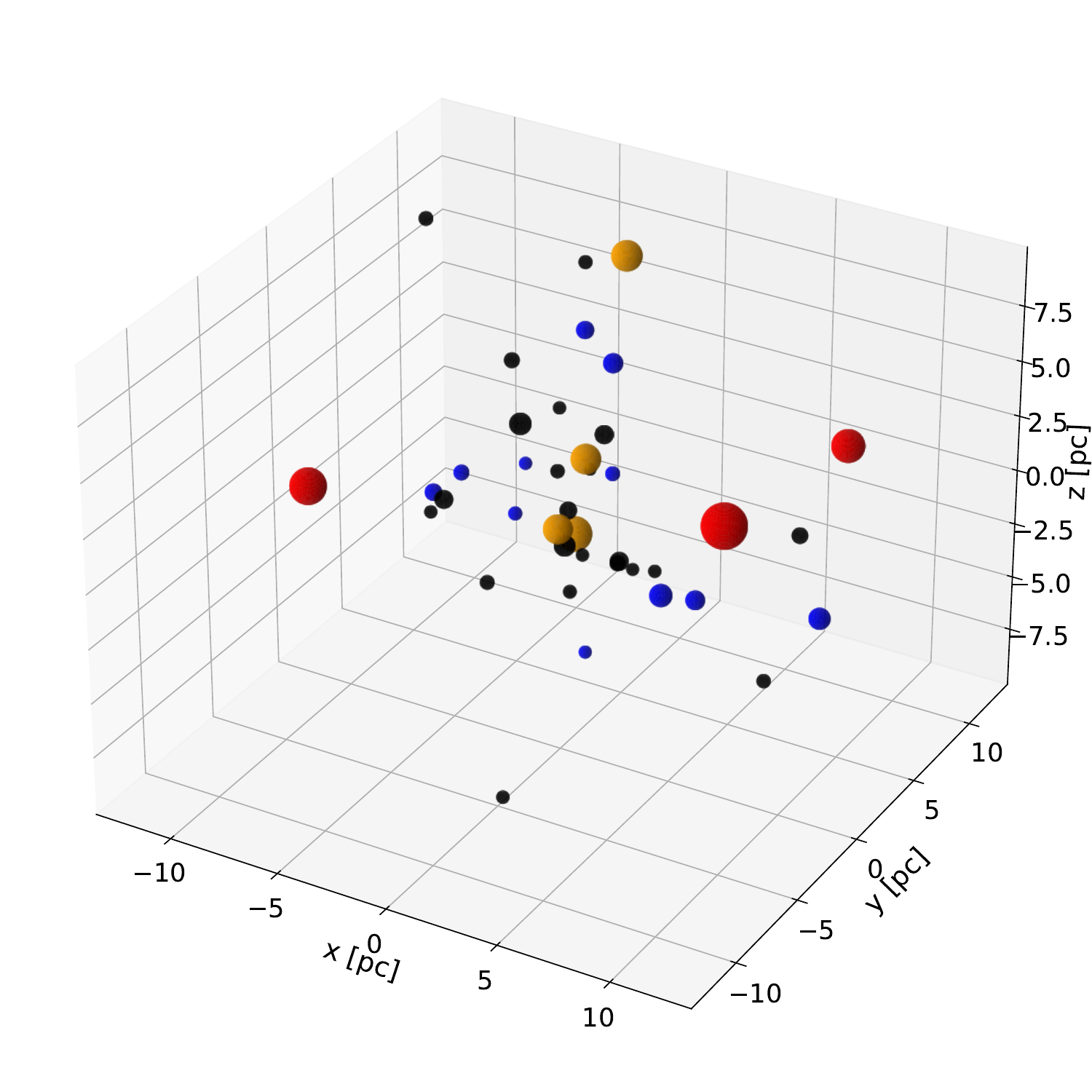}
   \caption{Simulated cluster. Red: Wolf-Rayet stars. Yellow: O stars above $10^{37}$~erg/s. Black: Remaining O stars from the 52 with inferred parameters. Blue: O stars from the synthetic population. The size of the spheres scales as $L_w^{1/5}$, with $L_w$ the mechanical power of the star.}
   \label{fig:clusterconfig3D}
 \end{figure}

The stars set-up in our simulation are listed in Table~\ref{fig:clusterconfig3D}. We stress that this only partly reproduces the observations, as measurement uncertainties on parallaxes and wind parameters called for some extrapolations which are described in Section~\ref{subsec:simulatedpopulation} and Section~\ref{subsec:3Dpositions}. Fig.~\ref{fig:clusterconfig3D} provides a 3D view of the simulated cluster.

\begin{table*}\label{tab:listsimulatedappendix}
\begin{tabular}{llllllll}
\bf $x$~[pc]      & \bf $y$~[pc]     & \bf $z$~[pc]      & \bf $\mdot$~[$M_\odot$/yr]    & \bf $\vinf$~[km/s] & \bf Method  & \bf Stellar type               &\bf  Comment     \\ \hline
6.67   & 11.89 & 0.00   & 3.24E-05 & 1440 & Inferred          & WR (after 1.6 Myr) & Y coordinate shifted from 14.5 pc to 11.9 pc \\
-10.63 & -1.93 & 0.00   & 1.26E-05 & 2900 & Inferred          & WR (after 1.6 Myr) & X coordinate shifted from 13.6 pc to 10.6 pc \\
6.48   & 1.61  & 1.29   & 2.40E-05 & 3500 & Inferred          & WR (after 1.6 Myr) &                                              \\
2.16   & 2.03  & -2.16  & 3.02E-06 & 400  & Inferred          & BHG                &   Cygnus~OB2~\#12                               \\
0.19   & 0.77  & -0.45  & 7.41E-06 & 3188 & Inferred          & O                  &    Cygnus~OB2~\#7                                          \\
1.64   & 2.32  & 11.66  & 6.34E-06 & 2403 & Inferred          & O                  &   Cygnus~OB2~\#11                                        \\
0.48   & 1.13  & 2.87   & 3.30E-06 & 3229 & Inferred          & O                  & Cygnus~OB2~\#22                                             \\
-0.06  & -0.10 & 0.13   & 4.82E-06 & 2541 & Inferred          & O                  & Merging of OB2~\#8B and OB2~\#8C                 \\
-1.03  & -1.39 & 5.25   & 1.28E-06 & 2352 & Inferred          & O                  &                                              \\
-2.45  & 7.76  & -0.03  & 5.92E-07 & 2372 & Inferred          & O                  &                                              \\
-1.19  & -6.96 & 4.54   & 3.93E-07 & 2637 & Inferred          & O                  &                                              \\
5.70   & -5.41 & 3.03   & 5.21E-07 & 2280 & Inferred          & O                  &                                              \\
0.06   & 0.48  & 0.74   & 4.16E-07 & 2331 & Inferred          & O                  &                                              \\
9.02   & 3.19  & 0.81   & 2.97E-07 & 2362 & Inferred          & O                  &                                              \\
2.84   & -0.42 & -0.45  & 2.24E-07 & 2474 & Inferred          & O                  &                                              \\
-2.80  & 0.97  & 6.48   & 2.18E-07 & 2474 & Inferred          & O                  &                                              \\
-2.58  & -1.42 & -2.42  & 1.79E-07 & 2352 & Inferred          & O                  &                                              \\
-7.60  & 2.48  & 10.89  & 1.51E-07 & 2484 & Inferred          & O                  &                                              \\
-0.13  & -0.06 & 2.74   & 2.35E-07 & 1862 & Inferred          & O                  &                                              \\
10.21  & -1.87 & -2.87  & 1.62E-07 & 2158 & Inferred          & O                  &                                              \\
0.10   & 0.55  & -3.06  & 1.27E-07 & 2352 & Inferred          & O                  &                                              \\
0.39   & 1.22  & 11.57  & 1.92E-07 & 1882 & Inferred          & O                  &                                              \\
-0.77  & -3.35 & -10.92 & 1.03E-07 & 2474 & Inferred          & O                  &                                              \\
-4.96  & -1.77 & 0.32   & 1.08E-07 & 2270 & Inferred          & O                  &                                              \\
4.48   & -0.35 & -0.35  & 9.54E-08 & 2372 & Inferred          & O                  &                                              \\
-0.03  & -0.06 & 5.61   & 9.41E-08 & 2352 & Inferred          & O                  &                                              \\
1.39   & -0.68 & -0.35  & 8.14E-08 & 2505 & Inferred          & O                  &                                              \\
0.48   & 1.51  & 2.19   & 1.31E-07 & 1791 & Inferred          & O                  &                                              \\
3.00   & 2.84  & -3.54  & 1.42E-06 & 2599 & IMF, $M = 48.5$ & O                  &                                              \\
10.31  & 2.58  & -2.32  & 1.04E-06 & 2580 & IMF, $M = 44.2$ & O                  &                                              \\
-2.32  & 8.31  & 3.06   & 7.70E-07 & 2563 & IMF, $M = 40.7$ & O                  &                                              \\
6.73   & -1.19 & -0.58  & 5.78E-07 & 2548 & IMF, $M = 37.7$ & O                  &                                              \\
-2.58  & 6.41  & 5.38   & 4.38E-07 & 2534 & IMF, $M = 35.2$ & O                  &                                              \\
-4.41  & -2.45 & 1.74   & 3.35E-07 & 2521 & IMF, $M = 33.1$ & O                  &                                              \\
-2.64  & -3.25 & 3.51   & 2.02E-07 & 2498 & IMF, $M = 29.6$ & O                  &                                              \\
1.48   & 1.55  & 2.29   & 1.59E-07 & 2488 & IMF, $M = 28.1$ & O                  &                                              \\
-2.93  & 1.39  & -0.71  & 1.26E-07 & 2479 & IMF, $M = 26.8$ & O                  &                                              \\
-0.77  & 3.32  & -7.60  & 1.01E-07 & 2470 & IMF, $M = 25.6$ & O                  &                                              \\
-0.39  & -2.13 & 3.99   & 8.14E-08 & 2462 & IMF, $M = 24.6$ & O                  &                                              \\
0.19   & 0.00  & -0.61  & 1.00E-06 & 2500 &  Fiducial                       & O                  & Putative SN progenitor             
\end{tabular}
\caption{List of the simulated stars. In the {\it Method} column, ``Inferred'' means that the wind parameters have been obtained from measured values of the wind-strength parameter and effective temperature. ``IMF'' means that the wind parameters have been generated from a synthetic population following an initial mass function of index 1.39, and we provide the initial mass in the Table. See Sections~\ref{subsec:simulatedpopulation},~\ref{subsec:3Dpositions} for details.}
\end{table*}

\bibliographystyle{mnras}
\bibliography{biblio} 






\bsp	
\label{lastpage}
\end{document}